\documentclass[12pt]{cernprep}
\usepackage{graphicx}
\usepackage{amssymb}
\usepackage{epsfig,color,rotating}
\begin{document}
\newcommand{\dedx}{\mbox{${\rm d}E/{\rm d}x$}}
\newcommand{\EcB}{$E \! \times \! B$}
\newcommand{\omt}{$\omega \tau$}
\newcommand{\omtsq}{$(\omega \tau )^2$}
\newcommand{\rphi}{\mbox{$r \! \cdot \! \phi$}}
\newcommand{\srphi}{\mbox{$\sigma_{r \! \cdot \! \phi}$}}
\newcommand{\dg}{\mbox{`durchgriff'}}
\newcommand{\mg}{\mbox{`margaritka'}}
\newcommand{\pT}{\mbox{$p_{\rm T}$}}
\newcommand{\GeVc}{\mbox{GeV/{\it c}}}
\newcommand{\MeVc}{\mbox{MeV/{\it c}}}
\def\kr{$^{83{\rm m}}$Kr\ }
\begin{titlepage}
%\docnum{CERN--PH--EP/2009--xxx}
\docnum{ }
\date{15 September 2009}
\vspace{1cm}
\title{{\Large Why the paper CERN--PH--EP--2009--015 (arXiv:0903.4762) 
\vspace{3mm} \\
is scientifically unacceptable}} 

\begin{abstract}
The paper CERN--PH--EP--2009--015 (arXiv:0903.4762) by A.~Bagulya {\it et al.} violates standards of quality of work and scientific ethics on several counts. The paper contains assertions that contradict established detector physics. The paper falls short of proving the correctness of the authors' concepts and results. The paper ignores or quotes misleadingly pertinent published work. The paper ignores the fact that the authors' concepts and results have already been shown wrong in the published literature. The authors seem unaware that cross-section results from the `HARP Collaboration' that are based on the paper's concepts and algorithms are in gross disagreement with the results of a second analysis of the same data, and with the results of other experiments. 
\end{abstract}

\vfill  \normalsize
\begin{center}
The HARP--CDP group  \\  

\vspace*{2mm} 

A.~Bolshakova$^1$, 
I.~Boyko$^1$, 
G.~Chelkov$^{1a}$, 
D.~Dedovitch$^1$, 
A.~Elagin$^{1b}$, 
M.~Gostkin$^1$,
A.~Guskov$^1$, 
Z.~Kroumchtein$^1$, 
Yu.~Nefedov$^1$, 
K.~Nikolaev$^1$, 
A.~Zhemchugov$^1$, 
F.~Dydak$^{2*}$, 
J.~Wotschack$^2*$, 
A.~De~Min$^{3c}$,
V.~Ammosov$^4$, 
V.~Gapienko$^4$, 
V.~Koreshev$^4$, 
A.~Semak$^4$, 
Yu.~Sviridov$^4$, 
E.~Usenko$^{4d}$, 
V.~Zaets$^4$ 
\\
 
\vspace*{5mm} 

$^1$~{\bf Joint Institute for Nuclear Research, Dubna, Russia} \\
$^2$~{\bf CERN, Geneva, Switzerland} \\ 
$^3$~{\bf Politecnico di Milano and INFN, 
Sezione di Milano-Bicocca, Milan, Italy} \\
$^4$~{\bf Institute of High Energy Physics, Protvino, Russia} \\

\vspace*{5mm}

%\submitted{(To be submitted to Eur. Phys. J. C)}
\end{center}

\vspace*{5mm}
\rule{0.9\textwidth}{0.2mm}

\begin{footnotesize}

$^a$~Also at the Moscow Institute of Physics and Technology, Moscow, Russia 

$^b$~Now at Texas A\&M University, College Station, USA 

$^c$~On leave of absence at 
Ecole Polytechnique F\'{e}d\'{e}rale, Lausanne, Switzerland 

$^d$~Now at Institute for Nuclear Research RAS, Moscow, Russia

$^*$~Corresponding author; e-mail: friedrich.dydak@cern.ch
\end{footnotesize}

\end{titlepage}

%\newpage
%\mbox{ }
%\tableofcontents
%\vspace{0.8cm}

\newpage 

\section{Prologue}

The `neutrino factory' (see Ref.~\cite{neutrinofactory} and further references cited therein) is a serious contender for a future accelerator facility that addresses fundamental questions on neutrino oscillations. One of the neutrino factory's many technological challenges is the production of charged pions with sufficient intensity to achieve the required particle fluxes in the decay chain pions $\rightarrow$ muons $\rightarrow$ neutrinos. 

The neutrino factory will cost several billion US dollars, {\it inter alia\/} for the requirement of a proton power on 
target at the 4~MW level to produce a sufficient number of charged pions.

With a view to the optimization of the design parameters of the proton driver of a neutrino factory, the HARP experiment was approved by CERN Management in 2000, and took data in 2001 and 2002. Beyond the primary aim of precise pion production data in the interactions of few GeV/{\it c} protons with heavy nuclei 
such as tantalum, the experiment was designed to deliver useful data for the understanding of the underlying physics and the modelling of Monte Carlo generators of hadron--nucleus collisions, for flux predictions of conventional neutrino beams, and for the calculation of the atmospheric neutrino flux.

A severe disagreement over concepts and quality of data analysis led to a split of the HARP Collaboration: on one side what continues to figure under `HARP Collaboration', aka `Official HARP' (OH), on the other side us, the `HARP--CDP' group. We performed our own analysis of the data, with nothing in common between our analysis and results and those of OH.  

Unfortunately, OH insisted on bringing pion production cross-sections into the public domain despite repeated warnings at various levels that their results are questionable.

Three `Comment' papers~\cite{NIMComment,IEEEComment,EPJCComment} as well as explicit analyses and proofs in CERN-internal papers (see Ref.~\cite{WhiteBookseries} and further references cited therein), in a refereed journal~\cite{JINSTpub}, and in a report to the CERN SPS and PS Experiments Committee 
(SPSC)~\cite{ReporttoSPSC} that OH's data analysis is seriously flawed, did not deter OH from disseminating their results. Out of many, we single out two papers that discuss and summarize the evidence that OH's analysis 
concepts and procedures are wrong:
\begin{enumerate}
\item `Comments on TPC and RPC calibrations reported by the HARP Collaboration'~\cite{JINSTpub}; and 
\item `On the flaws in ``Official'' HARP's data analysis'~\cite{ReporttoSPSC}.
\end{enumerate}

We stress that every single argument that OH brought forward in defence of their analysis concepts and procedures, has been shown to be not applicable or wrong in the above-cited publications. All that is published. There is no need for repetition here.

Unfortunately, there was no palpable effect of these publications on OH. Nor was there any effect from clear conclusions by independent review bodies: a HARP Review asked for by CERN and Italy's INFN~\cite{Foa}, and the 
CERN SPSC~\cite{CarliFuster,SPSCminutes}. Rather than accepting physics arguments, OH insist on ignoring unwelcome evidence.

The most recent example of this behaviour is their paper {\it `Dynamic Distortions in the HARP TPC: observations, measurements, modelling and corrections'}  by A.~Bagulya {\it et al.} that appeared recently as preprint CERN--PH--EP--2009--015 (arXiv:0903.4762). This paper violates standards of quality of work and scientific ethics. We cannot let it go on record without reaction.

\section{Pion production cross-sections}

\begin{figure}[ht]
\begin{center}
%\begin{tabular}{c}
\includegraphics[height=0.45\textwidth]{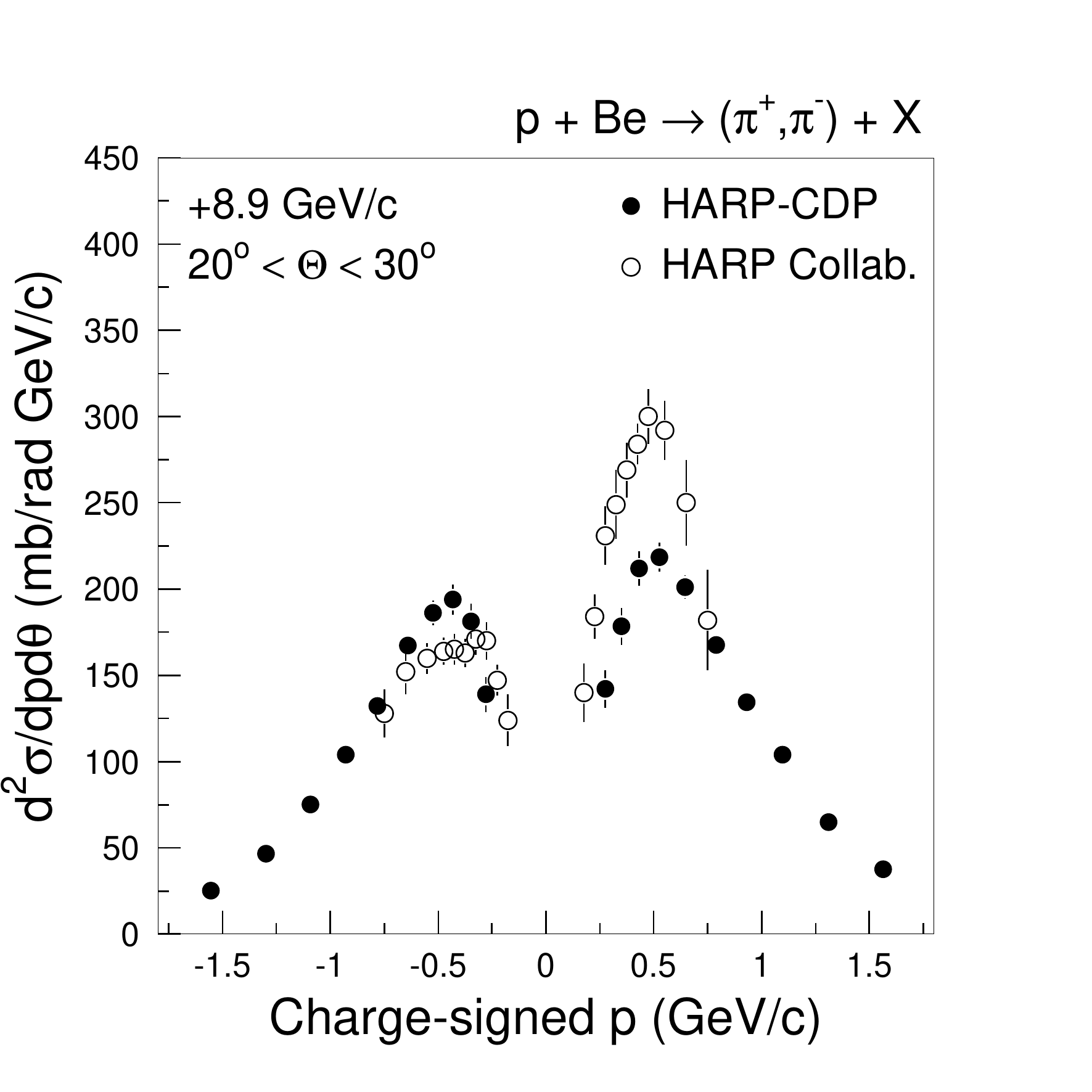} %\\ 
\includegraphics[height=0.45\textwidth]{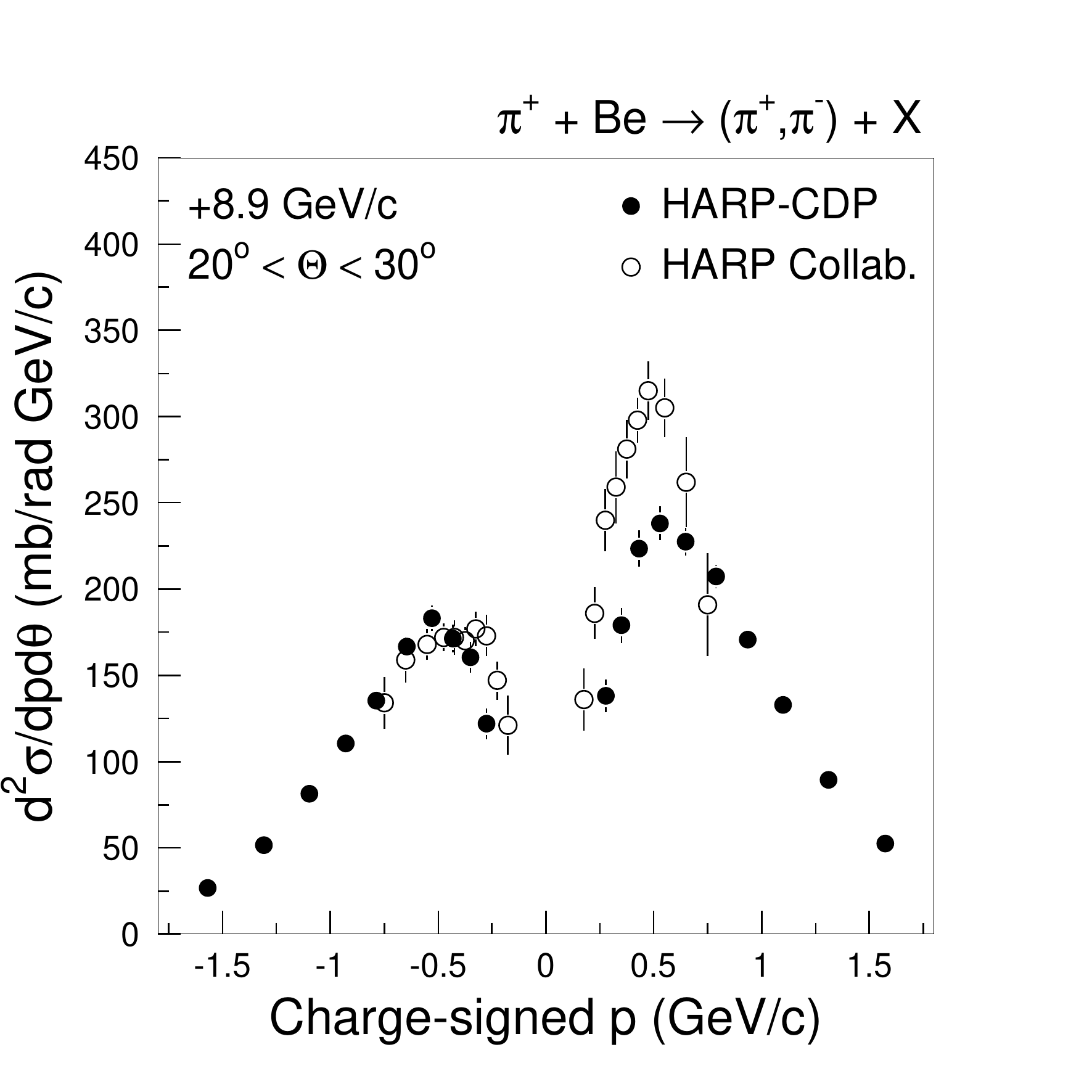} %\\
\includegraphics[height=0.45\textwidth]{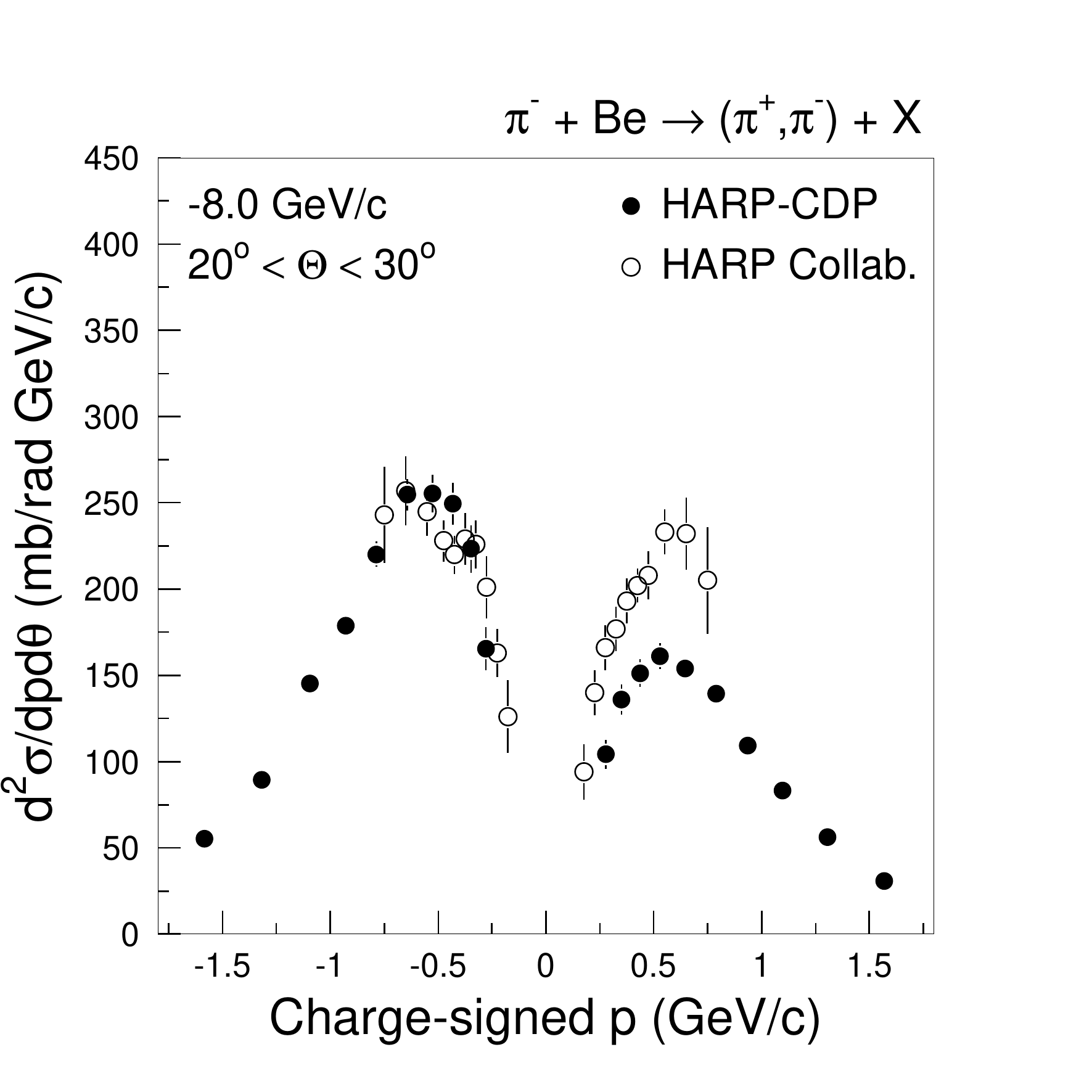} %\\
%\end{tabular}
\caption{Comparison of HARP--CDP cross-sections (black circles) of $\pi^\pm$ production by $+8.9$~GeV/{\it c} protons, $+8.9$~GeV/{\it c} $\pi^+$, and $-8.0$~GeV/{\it c} $\pi^-$, off beryllium nuclei, with the cross-sections 
published by the HARP Collaboration (open circles).} 
\label{BeComparisonWithOH}
\end{center}
\end{figure}

Figure~\ref{BeComparisonWithOH} shows the comparison of our cross-sections of $\pi^\pm$ production by $+8.9$~GeV/{\it c} protons, $+8.9$~GeV/{\it c} $\pi^+$, and $-8.0$~GeV/{\it c} $\pi^-$, off beryllium nuclei, with the ones published by OH~\cite{OffLAprotonpaper,OffLApionpaper}, in the polar-angle range $20^\circ < \theta < 30^\circ$. The latter cross-sections are plotted as published, while we expressed our cross-sections in the unit used by the HARP Collaboration. The errors shown are the published total errors.

\begin{figure}[ht]
\begin{center}
%\begin{tabular}{c}
\includegraphics[height=0.45\textwidth]{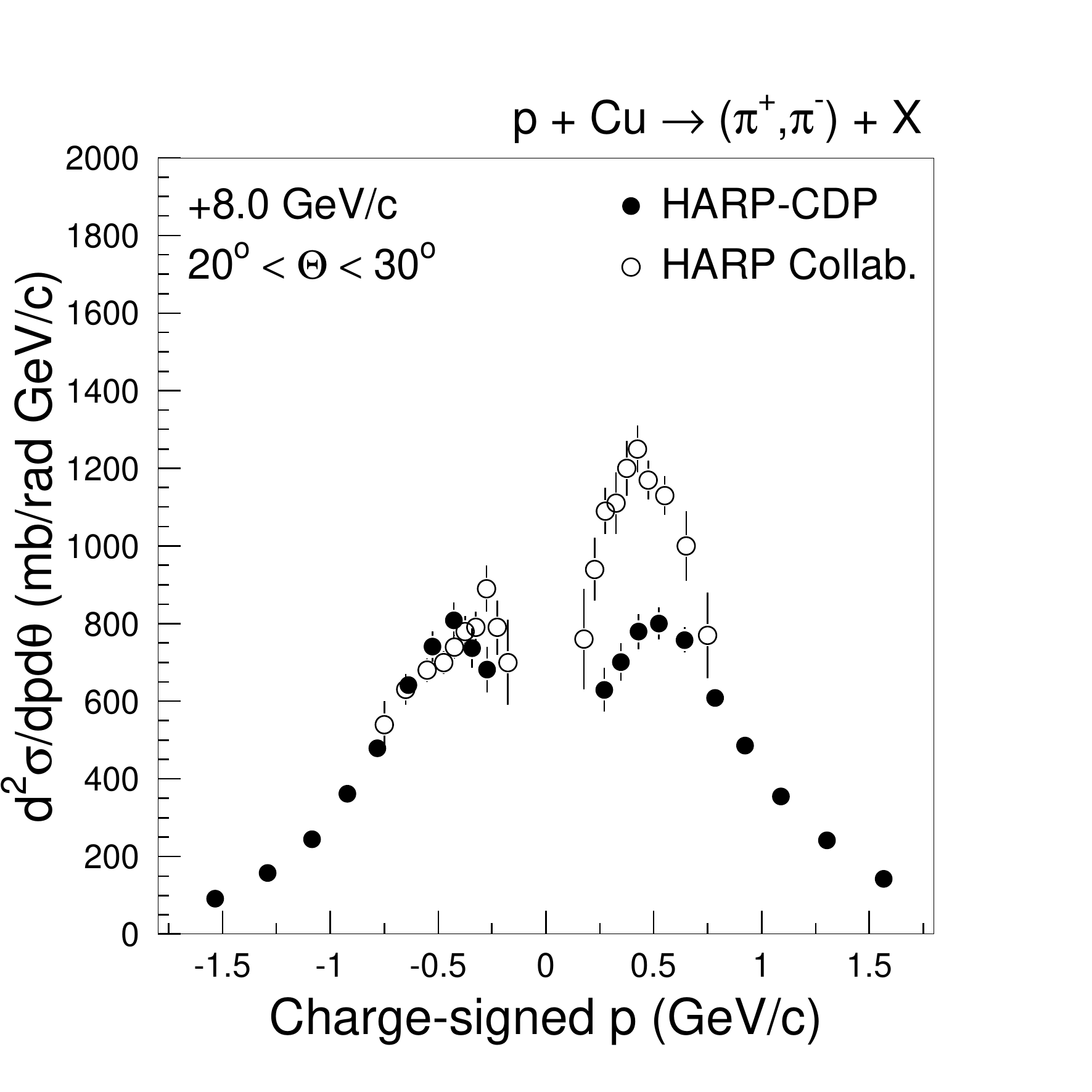} %\\ 
\includegraphics[height=0.45\textwidth]{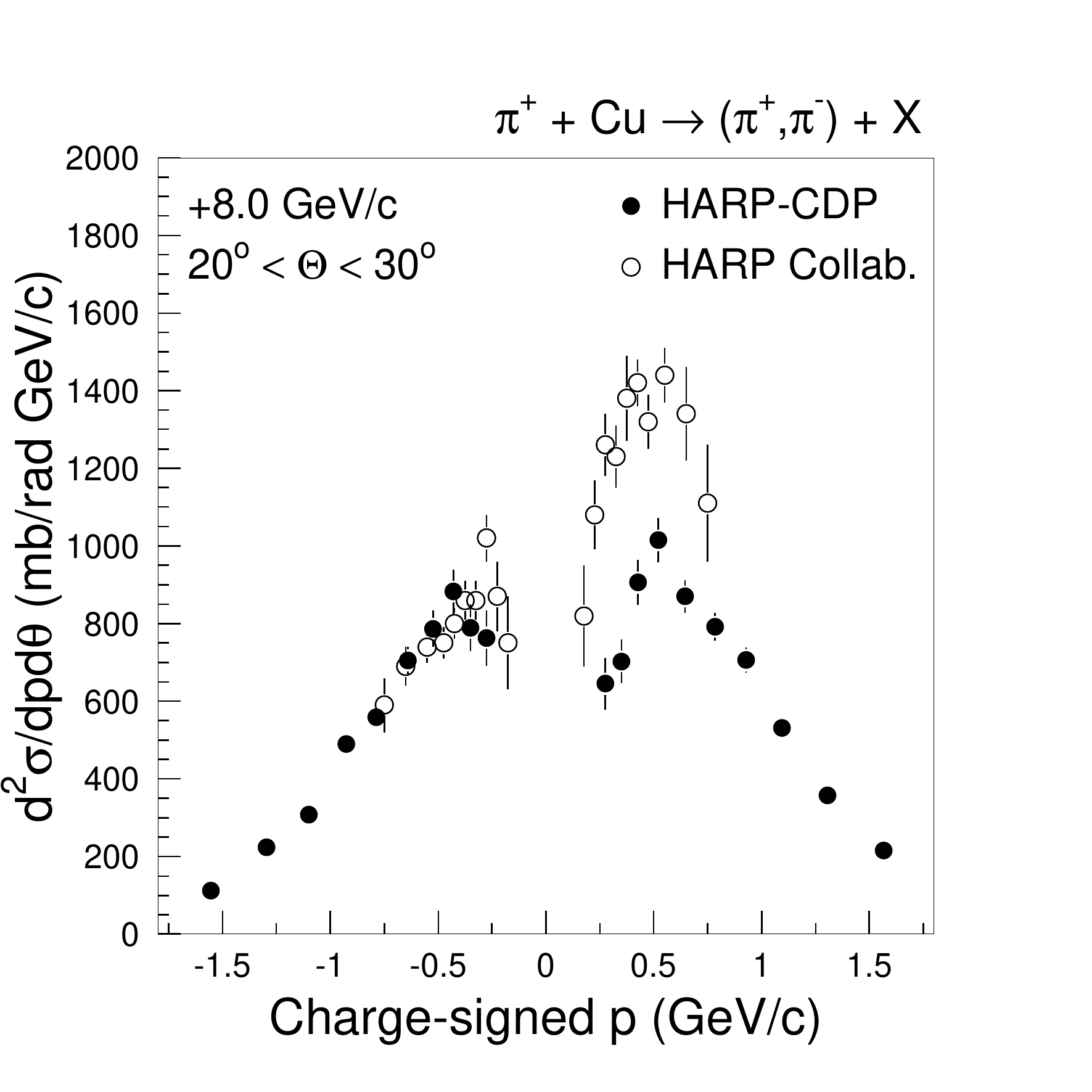} %\\
\includegraphics[height=0.45\textwidth]{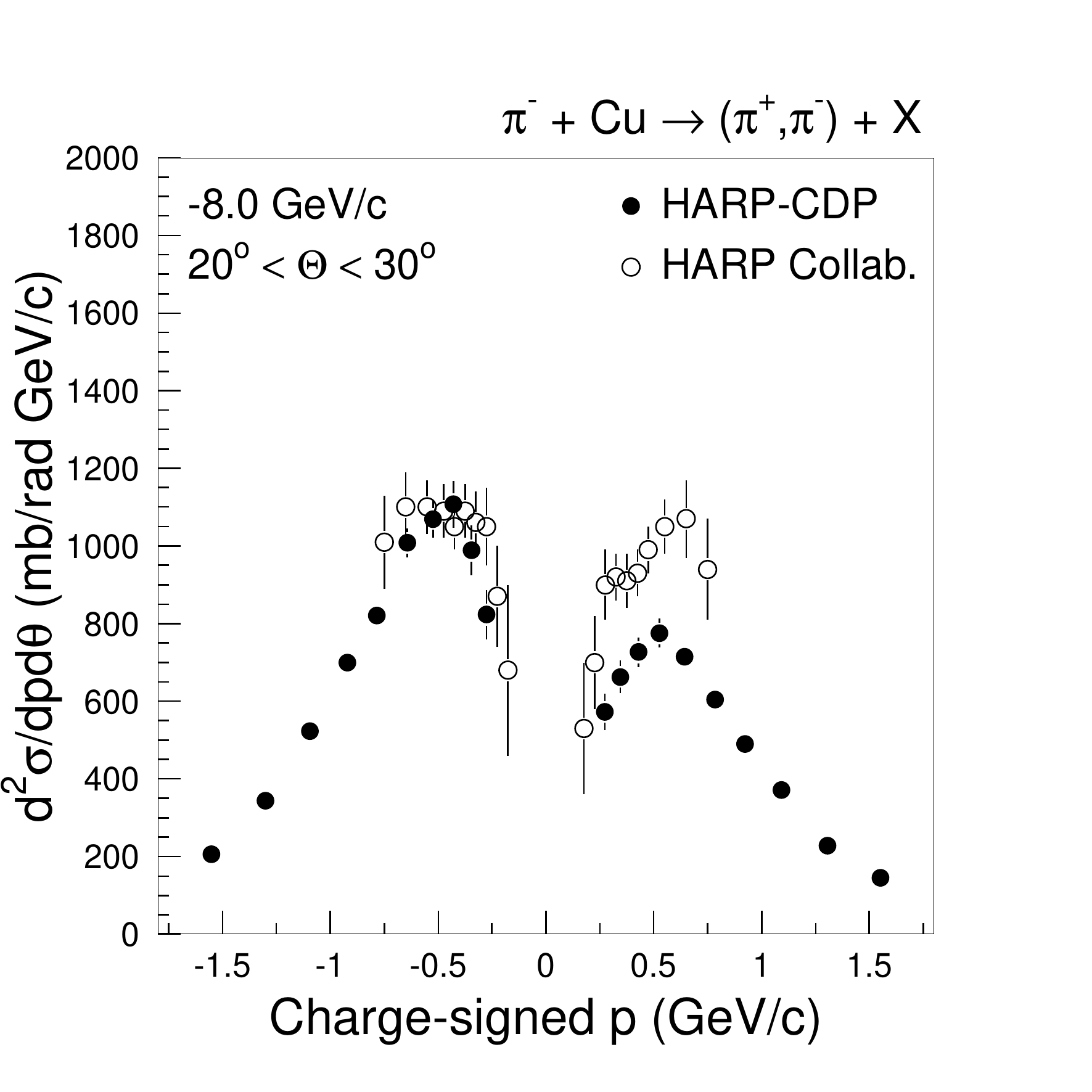} %\\
%\end{tabular}
\caption{Comparison of HARP--CDP cross-sections (black circles) of $\pi^\pm$ production by $+8.0$~GeV/{\it c} protons, $+8.0$~GeV/{\it c} $\pi^+$, and $-8.0$~GeV/{\it c} $\pi^-$, off copper nuclei, with the cross-sections 
published by the HARP Collaboration (open circles).} 
\label{CuComparisonWithOH}
\end{center}
\end{figure}

Figure~\ref{CuComparisonWithOH} shows the same comparison for copper nuclei, except that the beam momentum of protons and $\pi^+$ is not $+8.9$~GeV/{\it c} but $+8.0$~GeV/{\it c}.

\begin{figure}[h]
\begin{center}
%\begin{tabular}{c}
\includegraphics[height=0.45\textwidth]{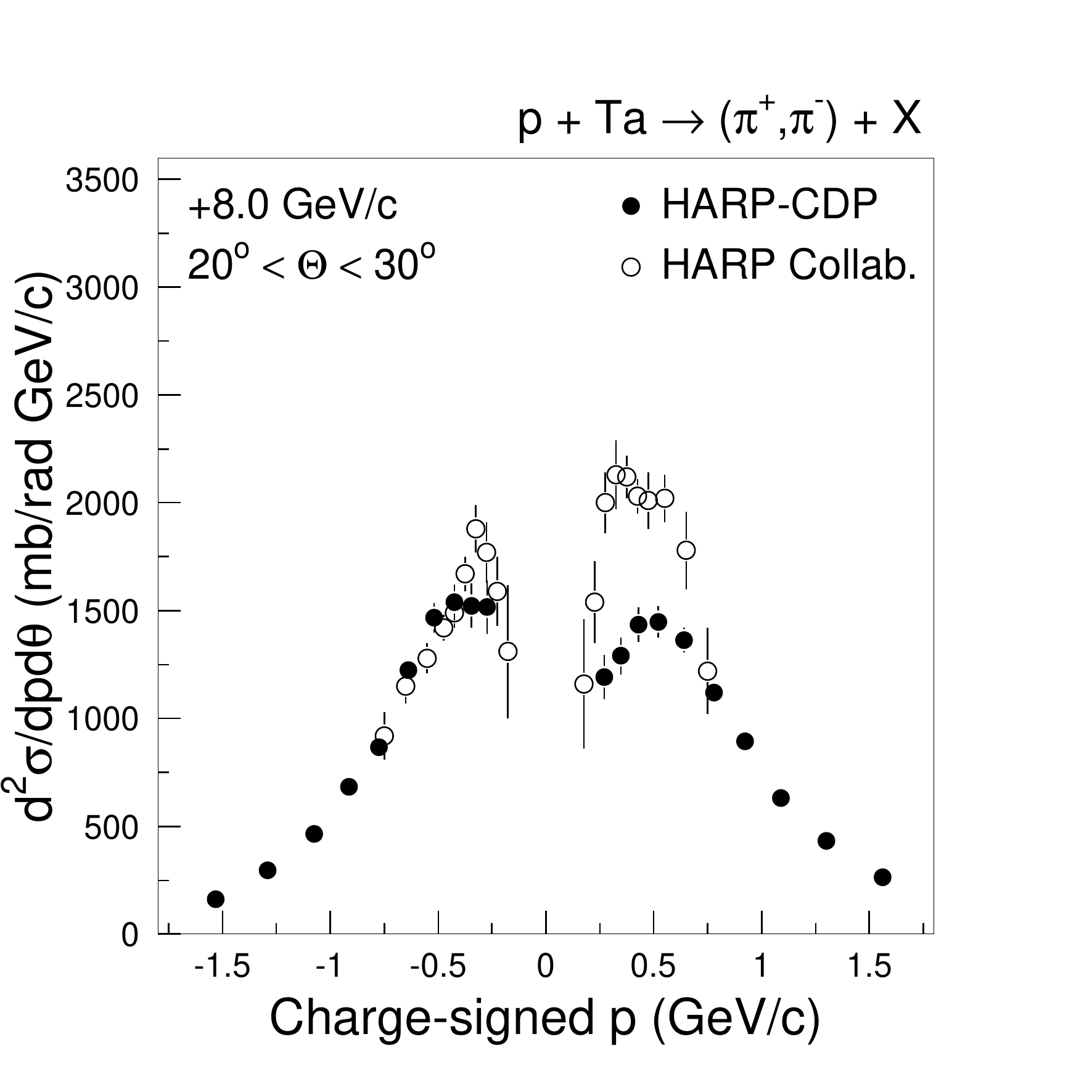} %\\ 
\includegraphics[height=0.45\textwidth]{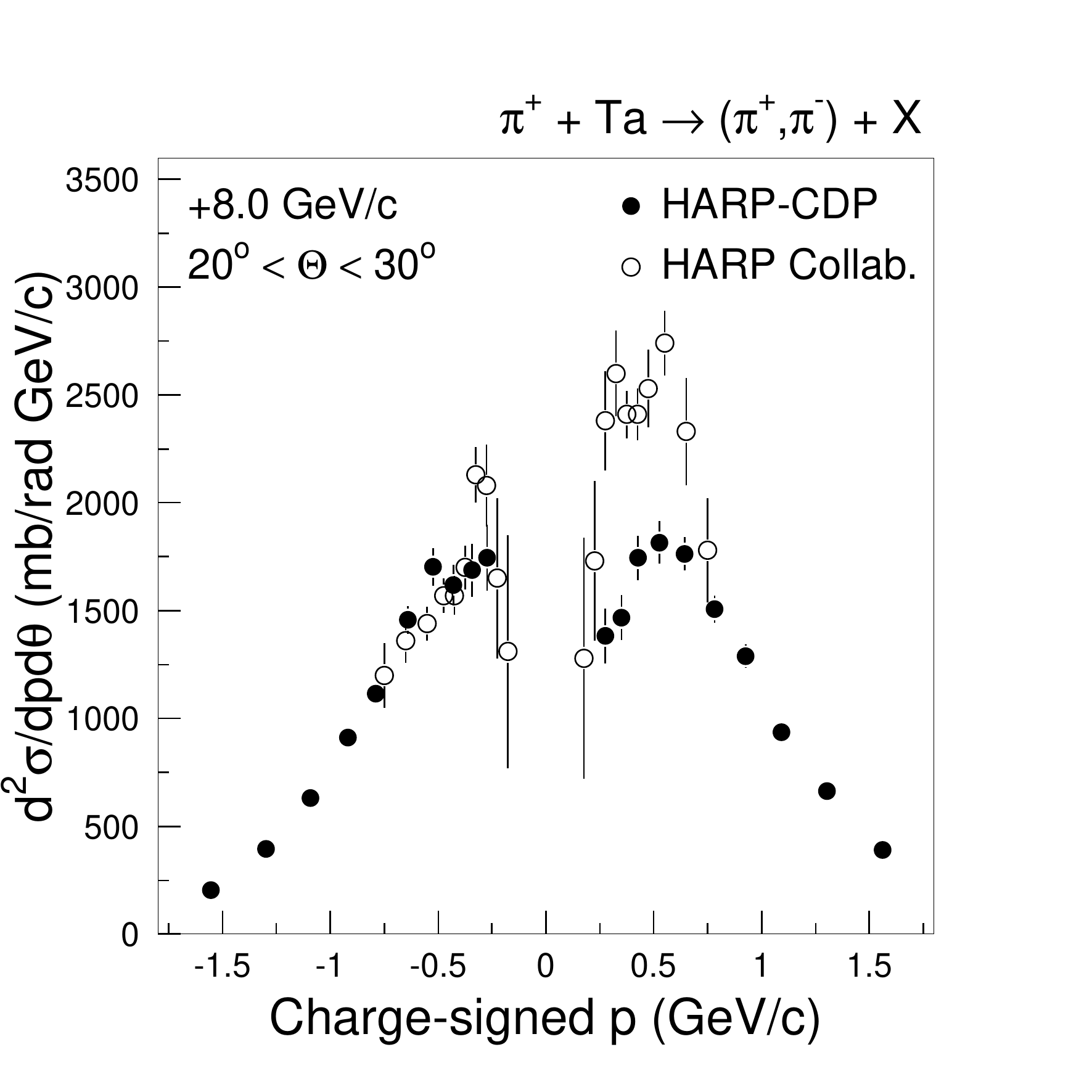} %\\
\includegraphics[height=0.45\textwidth]{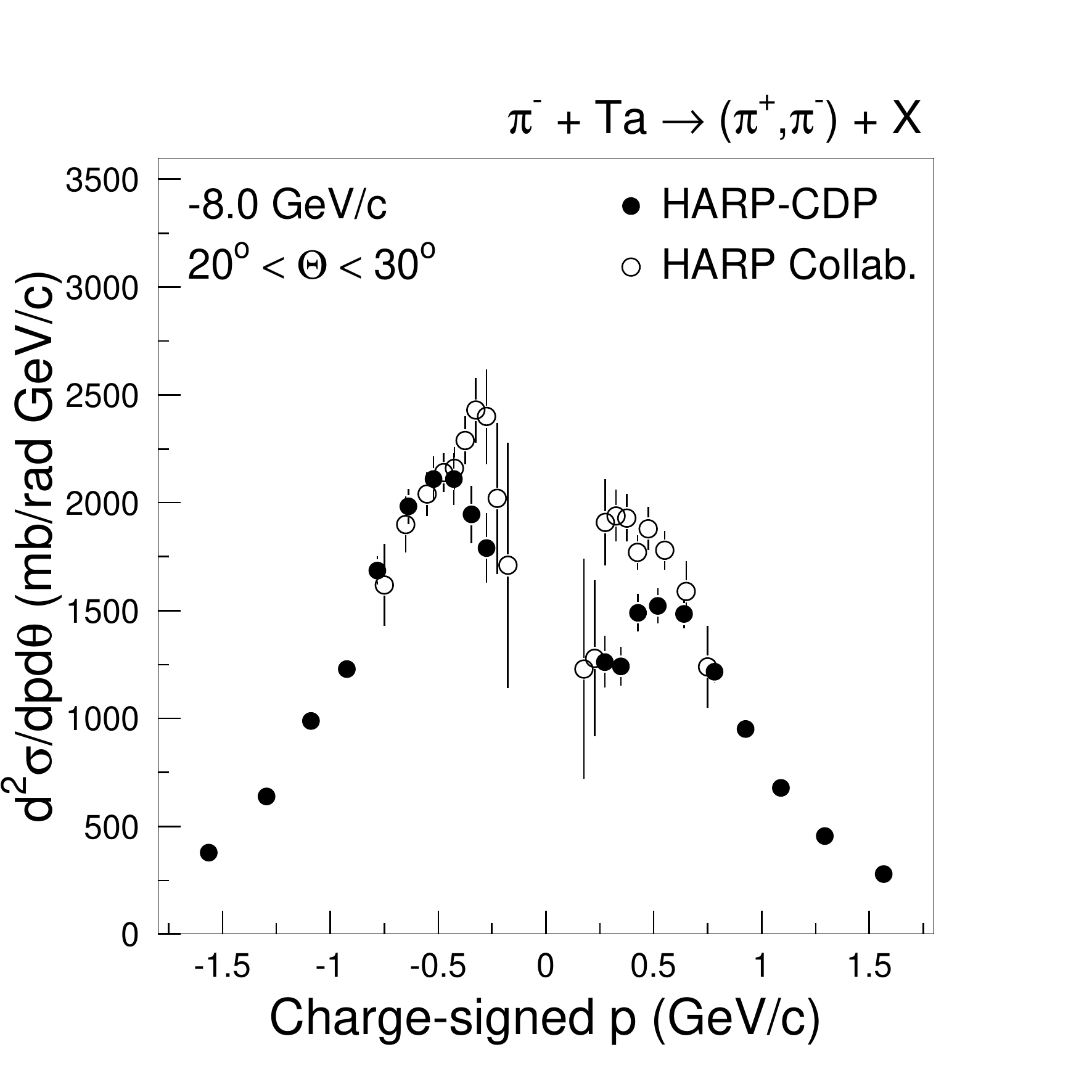} %\\
%\end{tabular}
\caption{Comparison of our cross-sections (black circles) of $\pi^\pm$ production by $+8.0$~GeV/{\it c} protons, $+8.0$~GeV/{\it c} $\pi^+$, and $-8.0$~GeV/{\it c} $\pi^-$, off tantalum nuclei, with the cross-sections published by the HARP Collaboration (open circles).} 
\label{TaComparisonWithOH}
\end{center}
\end{figure}

Figure~\ref{TaComparisonWithOH} shows the same comparison for tantalum nuclei.

The discrepancy between our results and those published by the HARP Collaboration qualifies in view of the quoted errors as `dramatic'. It is even more serious as the same data set has been analysed by both groups.

We note the dominant recurrent feature in the OH data that shows up in all $\pi^{\pm}$ production cross-sections irrespective of the target projectile and the target nucleus: a strong excess of positive pions. We also note that OH limit themselves to cross-sections for secondary pions with less than 0.8~GeV/{\it c} momentum, for which neither a physics reason nor an explanation is given.

In our papers we have presented overwhelming evidence of what went wrong in OH's analysis concepts and procedures. For a succinct summary we refer to the Appendix in Ref.~\cite{Bepub}.

OH's cross-sections not only disagree with our cross-sections, they also disagree with the results from the E802~\cite{E802} and E910~\cite{E910} experiments (see also Refs.~\cite{Bepub,Bepub2,Tapub,Cupub} for the comparison of cross-sections.)

We note that OH do not utter a single word about these quite remarkable disagreements.

\section{Mistakes, circular argumentations, and contradictions of accepted detector physics}

Leaving aside that we argued and showed explicitly that OH's understanding of the track distortions in the HARP TPC is flawed, and that their simple-minded procedures fall way short of the required accuracy of correction, in a physics paper one expects that claims are not only made one after the other, but also shown to hold. This is not the case in the paper in question. In the sequence of claims, none of them is shown to be valid. Rather, exactly the same claims are made as before in Ref.~\cite{OffTPCcalibration}, with no regard whatsoever to the fact that all these claims have been explicitly and numerically shown to be wrong in Ref.~\cite{JINSTpub}.

We point to two particularly telling instances.

First, OH take their $d_0^\prime$ parameter\footnote{$d_0^\prime$ is the signed impact parameter of a track in the transverse $x$-$y$ plane, determined from a circle fit of the distorted TPC cluster positions of a track.} as a measure of TPC track distortions, and adjust the amplitude of their distortion corrections so as to minimize $d_0^\prime$. The parameter $d_0^\prime$, however, is unsuitable for this purpose:
\begin{itemize}
\item the use of $d_0^\prime$ underestimates the distortions because the fitted track `co-moves' with the distorted cluster positions, and because the fit measures only the relative deviation from a circle (while the actual TPC distortions seriously distort the theoretical circle in a perfect solenoidal magnetic field); 
\item $d_0^\prime$ lumps together the TPC distortions into one single number and thus misses out on the complicated radial and $z$ dependences of distortions.
\end{itemize}

In other words: from the outset, an unsuitable parameter is used to determine a correction; applying this correction then means in no way that the distortion is eliminated. It is obvious that the correct size of distortions can be determined solely by reference to an external system that is not affected by TPC track distortions\footnote{For this purpose, we have demonstrated the successful use of the geometrical positions of the RPCs that surround the TPC; our procedure and results, including the stunning differences from the results achieved when not using the RPCs, are published, for example, in Ref.~\cite{TPCpub}.}.

\clearpage

The second example is their claim that their cross-section results agree when determined without any distortion correction from data at the start of the spill\footnote{OH claim---incorrectly, as shown by us e.g. in Ref.~\cite{TPCpub}---that there are no TPC track distortions at the start of the spill, and therefore the data taken during the first $\sim$25\% of the spill need no distortion corrections.}, and when determined from all data in
the spill after distortion corrections.

Here, OH avoid telling explicitly that their procedure forces the remainder of the spill to reproduce the start of the
spill. We recall that at the start of the spill, OH consider their parameter $d_0^\prime$ sufficiently close to zero and use this as argument that no correction for TPC track distortions is necessary (an unbiased observer of their plots would, however, not easily reproduce this conclusion, see, e.g., the $d_0^\prime$ dependence as a function of the time in spill in Ref.~\cite{OffTapaper}). In the remainder of the spill, corrections are applied so as to force $d_0^\prime$ as much as possible to zero. Therefore, whatever error is already in their cross-sections at the start of the spill, it is reproduced in the remainder of the spill. The circularity of the procedure is obvious. To make things worse: OH do not stop short of claiming improved precision from using the data from the full spill.  

As for contradictions of accepted detector physics, we point to two particularly telling instances.

In Section 5, the paper estimates 25~ms as the time that Ar$^+$ ions need to drift a distance of 11~mm. Since the respective electric field strength is some 1600~V/cm, it follows from the known velocity of Ar$^+$ ions in gas that the respective time is one order of magnitude smaller than 25~ms. Rather than regarding this discrepancy as a reason to call their concept of quantifying the size of track distortion into question, they argue with their observation---without any regard to the fact that it cannot possibly be correct. 

Not explicitly stated, but silently underlying the whole paper, is their claim that the timing of protons in the HARP RPCs is advanced by some 500~ps, a feature that they published as a novel detector effect in fast-timing RPCs~\cite{500pseffect}. Needless to say, fast-timing RPCs are well enough understood that this effect can be
safely excluded. What is not understandable is that the obvious reason for their `discovery', namely a bias in the proton momentum of the size $\Delta (1/p_{\rm T}) \simeq 0.3$~(GeV/{\it c})$^{-1}$ in their reconstruction of TPC tracks, caused by a lack of understanding of TPC track distortions, is discarded as the source of their 500~ps effect. Rather than trying to understand and remove this source of error, they chose to claim a novel detector physics effect.

\section{Violations of scientific ethics}

Scientific ethics require that if there is earlier published work by others that calls into question the current work,
the respective arguments must be referred to and discussed. If one insists on one's point of view, this is only permitted if one explicitly discusses why the arguments made earlier are not applicable or are wrong.

OH have never been able to counter our arguments, nor those made by independent review committees, or show them not applicable or wrong. Rather, they stick to the policy of ignoring criticism, apparently in the hope of pulling the wool over the eyes of an uncritical public.

\section{Epilogue}

We trust that this reply to a paper that has reached the public domain, even though it should have been stopped by internal review, helps toward a more scientific rigour in the publication of physics results.

%\clearpage

\end{document}